\documentclass[twocolumn,prb,showpacs]{revtex4}

\usepackage{graphicx} \usepackage{amssymb} \usepackage{rotating}

\newcommand{\la}{\mathord{\langle}}
\newcommand{\ra}{\mathord{\rangle}}

\begin{document}

\title{In-plane   magnetic  reorientation   in   coupled  ferro-   and
antiferromagnetic thin films}

\author{P.J.  Jensen}  \altaffiliation[On  leave from  the  ]{Institut
f\"ur Theoretische Physik,  Freie Universit\"at Berlin, Arnimallee 14,
D-14195       Berlin,       Germany;       jensen@physik.fu-berlin.de}
\author{H. Dreyss\'e}

\affiliation{IPCMS --  GEMME, Universit\'e  Louis Pasteur, 23,  rue du
Loess, F-67037 Strasbourg, France }

\begin{abstract} 
By studying coupled ferro-  (FM) and antiferromagnetic (AFM) thin film
systems, we obtain an in-plane magnetic reorientation as a function of
temperature and  FM film thickness.  The  interlayer exchange coupling
causes  a uniaxial anisotropy,  which may  compete with  the intrinsic
anisotropy of the FM film.  Depending on the latter the total in-plane
anisotropy of the FM film  is either enhanced or reduced. Eventually a
change of sign occurs, resulting in an in-plane magnetic reorientation
between a collinear and an  orthogonal magnetic arrangement of the two
subsystems. A canted magnetic arrangement may occur, mediating between
these two  extremes. By measuring  the anisotropy below and  above the
N\'eel temperature the interlayer exchange coupling can be determined.
The   calculations  have   been  performed   with   a  Heisenberg-like
Hamiltonian by application of a two-spin mean-field theory.
\end{abstract} 
 \maketitle

The interface between coupled  ferro- (FM) and antiferromagnetic (AFM)
films  or particles  has  been attracted  much  interest recently,  in
particular due to the renewed  interest in exchange biased systems for
application   in    magnetoresistive   sensors.\cite{NoS99,Kiw01}   Of
particular interest is the case  of a 'compensated' AFM interface with
an equal number of  positive and negative exchange interactions across
the  interface. By considering  only exchange  couplings, it  has been
shown by N.C.\ Koon that the most stable magnetic arrangement for such
an interface is  an orthogonal magnetic orientation of  the FM and AFM
subsystems.\cite{Koo97}  A  nonvanishing  magnetic binding  energy  is
obtained if  the magnetic  moments of the  AFM are allowed  to deviate
from their equilibrium AFM arrangement, exhibiting thus a noncollinear
AFM magnetization  with a small component parallel  or antiparallel to
the FM,  the 'spin-flop-phase'.\cite{HiM86} From the  viewpoint of the
FM  film,  the net  magnetic  binding  energy  introduces an  in-plane
uniaxial  magnetic anisotropy  $K_\mathrm{int}$.\cite{ScB98}  A simple
estimate\cite{rem1} yields  the strength of  this interface anisotropy
to              be              of              the              order
$K_\mathrm{int}\propto-(J_\mathrm{int})^2/|J_\mathrm{AFM}|$,       with
$J_\mathrm{int}$ the interlayer  exchange coupling between neighboring
FM  and  AFM spins  across  the  interface,  and $J_\mathrm{AFM}$  the
exchange  coupling in  the AFM  system.  Experimentally,  collinear as
well  as orthogonal  magnetic arrangements  of coupled  FM-AFM systems
have been observed.\cite{NoS99,MNL98}

Evidently, the magnetic  direction of the FM film  depends also on its
intrinsic   anisotropy  $K_\mathrm{FM}$.    If   $K_\mathrm{int}$  and
$K_\mathrm{FM}$  favor   the  same  in-plane  easy   axis,  the  total
anisotropy   is   enhanced.\cite{Jen01}   If   the   two   anisotropic
contributions  favor   different  magnetic  directions,   an  in-plane
magnetic  reorientation  may  occur  as  a function  of  the  FM  film
thickness,  since $K_\mathrm{int}$  is proportional  to  the interface
area, whereas $K_\mathrm{FM}$ is porportional  to the volume of the FM
film.

Quite interestingly,  also a  different temperature behavior  of these
two  competing  anisotropies  may   result  in  an  in-plane  magnetic
reorientation. At  finite temperatures  $T$ the magnetic  direction is
determined   by    effective,   temperature   dependent   anisotropies
$\mathcal{K}(T)$,  which depend  on  $T$ mainly  through the  relative
magnetization   $M(T)$   as   can   be   shown   by   a   perturbative
treatment.\cite{CaC66,JeB98} In  the present case of  a coupled FM-AFM
system   the    main   reason   for   the    different   behavior   of
$\mathcal{K}_\mathrm{FM}(T)$ and  $\mathcal{K}_\mathrm{int}(T)$ is the
different ordering  temperature of the  two subsystems.  If  the Curie
temperature $T_C$ of the FM film is larger than the N\'eel temperature
$T_N$ of  the AFM system, then  for $T>T_N$ the  magnetic direction of
the FM film is exclusively determined by $\mathcal{K}_\mathrm{FM}(T)$,
whereas  below   $T_N$  it  depends   on  the  relative   strength  of
$\mathcal{K}_\mathrm{FM}(T)$     and    $\mathcal{K}_\mathrm{int}(T)$.
Hence,     the     total     anisotropy     of     the     FM     film
$\mathcal{K}_\mathrm{tot,FM}(T)=
\mathcal{K}_\mathrm{FM}(T)+\mathcal{K}_\mathrm{int}(T)$        possibly
exhibits  a  change of  sign  as a  function  of  temperature, and  an
in-plane magnetic reorientation occurs.

To our knowledge, such a magnetic reorientation in coupled FM-AFM thin
film systems has not been reported  yet.  In the present study we will
investigate  this phenomenon by  determining the  magnetic arrangement
and        the        temperature        dependent        anisotropies
$\mathcal{K}_\mathrm{FM}(T)$   and  $\mathcal{K}_\mathrm{int}(T)$.   A
Heisenberg-like  Hamilton operator is  applied with  localized quantum
spins \textbf{S}$_i$ and  spin quantum number $S=1$ on  a simple cubic
(001) lattice:
\begin{equation} 
\mathcal{H}=-\frac{1}{2}\sum_{\la       i,j\ra}\,J_{ij}\,\mathbf{S}_i\;
\mathbf{S}_j\; -\sum_i\,K_i\,(S_i^z)^2 \,. \label{e1}
\end{equation}
The FM  and AFM  films are assumed  to consist of  $n_\mathrm{FM}$ and
$n_\mathrm{AFM}$  atomic   layers,  spanned  by   the  $xz$-plane.   A
compensated  AFM interface is  considered, which  is accounted  for by
using two sublattices per layer.\cite{noncom} The exchange interaction
$J_{ij}$  couples nearest  neighbor  spins on  lattice  sites $i$  and
$j$.  Caused  by  the  shape  anisotropy  resulting  from  the  dipole
interaction, the magnetizations $\mathbf{M}_i=\la \mathbf{S}_i\ra$ are
confined to the film  plane.  Furthermore, we assume a layer-dependent
second order in-plane uniaxial  anisotropy $K_i$, favoring for $K_i>0$
an   easy  axis   along   the   $z$-  and   for   $K_i<0$  along   the
$x$-direction.\cite{rem2} The FM  and AFM subsystems are characterized
by the exchange couplings $J_\mathrm{FM}$ and $J_\mathrm{AFM}$, and by
the intrinsic  anisotropies $K_\mathrm{FM}$ and  $K_\mathrm{AFM}$. For
these quantities typical values  are taken into account. An anisotropy
for the  AFM is required, since  otherwise it will start  to rotate in
accordance  with the  FM film.   We  do not  distinguish here  between
surface  or interface anisotropies  different from  those of  the film
interior layers, although  they might differ considerably.\cite{HeB94}
FM and AFM are coupled across the interface by the interlayer exchange
coupling $J_\mathrm{int}$.

The site-dependent in-plane  magnetizations $\mathbf{M}_i(T)$ and free
energies $F_i(T)$ are calculated within  a mean field theory.  To take
into account  at least partly the  strong correlations in  the AFM, we
apply here a  two-spin-cluster (Oguchi-) method,\cite{Ogu55} with both
spins located in the same  layer.  Within this method the interactions
in the  cluster are treated  exactly, whereas the remaining  system is
considered by  a molecular field.   The free energies  and expectation
values  are  determined by  diagonalizing  the corresponding  two-spin
matrices.  We emphasize  that the  effective anisotropies  are neither
approximated     by    the     low-temperature    estimate\cite{CaC66}
$\mathcal{K}(T)\sim M^{l(l+1)}(T)$,  $l$ the order  of the anisotropy,
nor  by a thermodynamic  perturbation theory.\cite{JeB98}  This allows
for an appropriate  treatment of the anisotropies also  near and above
ordering temperatures.

Caused  by $J_\mathrm{int}$  the spins  of the  AFM  layers especially
close to the interface  may deviate from their undisturbed equilibrium
directions. For simplicity, due  to the strong FM exchange interaction
a collinear magnetization of the  whole FM film is assumed, which will
be   rotated   by   the   in-plane  angle   $\phi_\mathrm{FM}$.    The
magnetizations $|\mathbf{M}_\mathrm{i,FM}|(T)$  of the FM  layers, and
the   two   magnetization   components   $M_\mathrm{i,AFM}^x(T)$   and
$M_\mathrm{i,AFM}^z(T)$ of the AFM layers are determined by minimizing
the                  total                 free                 energy
$F(T,\phi_\mathrm{FM})=\sum_iF_i(T,\phi_\mathrm{FM})$ with the help of
a conjugated  gradient method. The  minimum of $F(T,\phi_\mathrm{FM})$
yields  the  equilibrium angle  $\phi_\mathrm{0,FM}$  of  the FM  film
magnetization.  The total  anisotropy $\mathcal{K}_\mathrm{tot,FM}(T)$
per FM spin is calculated  from the free energy difference between the
orthogonal     ($\phi_\mathrm{FM}=\pi/2$)     and    the     collinear
($\phi_\mathrm{FM}=0$) magnetic arrangement:
\begin{equation} 
\mathcal{K}_\mathrm{tot,FM}(T)=\frac{1}{n_\mathrm{FM}}     \,    \Big[
F(T,\phi_\mathrm{FM}=\pi/2)-       F(T,\phi_\mathrm{FM}=0)       \Big]
\,. \label{e2} \end{equation}

The  following results are  calculated assuming  representative values
for   the   exchange   and   anisotropy   parameters   in   units   of
$J_\mathrm{FM}$.      If    not     stated    otherwise,     we    use
$J_\mathrm{AFM}/J_\mathrm{FM}=-0.5$                                 and
$|K_\mathrm{FM}/J_\mathrm{FM}|=K_\mathrm{AFM}/J_\mathrm{FM}=0.01$.
For   the  thicknesses   of   the   FM  and   AFM   films  we   assume
$n_\mathrm{FM}=5$  and  $n_\mathrm{AFM}=10$.  From  these  values  the
critical        temperatures        $T_C/J_\mathrm{FM}=3.68$       and
$T_N/J_\mathrm{FM}=1.92$ are obtained for $J_\mathrm{int}=0$.
\begin{figure} 
\includegraphics[width=5.5cm,height=7.5cm,bb=70   30   590   700,clip,
angle=-90]{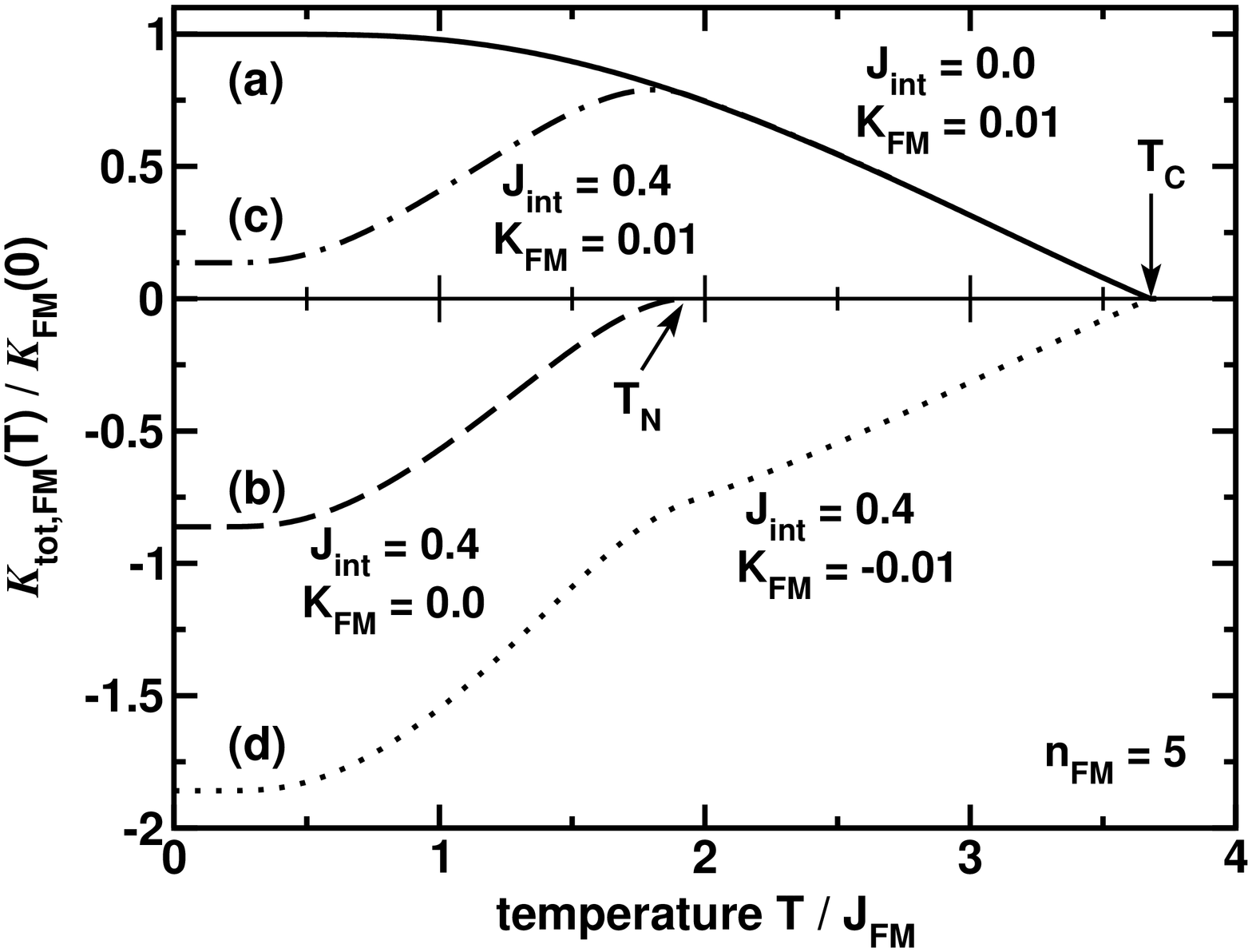}
\caption{ 
Total  effective anisotropy $\mathcal{K}_\mathrm{tot,FM}(T)$  per spin
of    the   FM   film    as   a    function   of    temperature   $T$.
$\mathcal{K}_\mathrm{tot,FM}(T)>0$    prefers    a   collinear,    and
$\mathcal{K}_\mathrm{tot,FM}(T)<0$ an orthogonal FM film magnetization
with respect  to the AFM  magnetic direction. The  exchange couplings,
intrinsic  anisotropies,  and  temperatures  are  given  in  units  of
$J_\mathrm{FM}$,  and  $\mathcal{K}_\mathrm{tot,FM}(T)$  in  units  of
$K_\mathrm{FM}=\mathcal{K}_\mathrm{FM}(T=0)$.         We        assume
$J_\mathrm{AFM}/J_\mathrm{FM}=-0.5$                                 and
$K_\mathrm{AFM}/J_\mathrm{FM}=0.01$, in addition $n_\mathrm{FM}=5$ and
$n_\mathrm{AFM}=10$ for the  thicknesses of the FM and  AFM films. For
these values the Curie temperature $T_C$ of the FM film is larger than
the N\'eel temperature $T_N$ of the  AFM film. The full line (a) shows
the intrinsic anisotropy for  a decoupled FM film ($J_\mathrm{int}=0$)
for $K_\mathrm{FM}/J_\mathrm{FM}=0.01$. The  dashed line (b) refers to
the   bare    interface   anisotropy   ($K_\mathrm{FM}=0$),   assuming
$J_\mathrm{int}/J_\mathrm{FM}=0.4$.  The presence of  both anisotropic
contributions  results in a  reduced $\mathcal{K}_\mathrm{tot,FM}(T)$,
dot-dashed   line  (c).  For   $K_\mathrm{FM}/J_\mathrm{FM}=-0.01$  an
enhanced   absolute    value   $|\mathcal{K}_\mathrm{tot,FM}(T)|$   is
obtained, dotted line (d). } \vspace*{-0.5cm} \end{figure}

In  Fig.1 we  show the  total effective  anisotropy per  FM spin  as a
function of  temperature $T$, where we depict  different scenarios.  A
positive value  of $\mathcal{K}_\mathrm{tot,FM}(T)$ favors  a magnetic
direction  of the  FM film  along the  $z$-axis collinear  to  the AFM
magnetization  (collinear   arrangement),  and  a   negative  value  a
direction along the $x$-axis (orthogonal arrangement).  The solid line
(a) refers to the intrinsic anisotropy $\mathcal{K}_\mathrm{FM}(T)$ of
the       FM       film       for       the       uncoupled       case
($J_\mathrm{int}=0$).   $\mathcal{K}_\mathrm{FM}(T)$   decreases  with
increasing  temperature   and  vanishes  for  $T>T_C$,   as  has  been
calculated   and   measured   for   many  different   FM   thin   film
systems.\cite{HeB94}   This  does  not   imply  that   the  underlying
spin-orbit  coupling  varies  with  temperature.  Rather  due  to  the
increasing thermal agitation the ability of the anisotropy to maintain
a  particular direction  of the  magnetization decreases.   The dashed
line (b) shows  the interface anisotropy $\mathcal{K}_\mathrm{int}(T)$
for     a    vanishing     intrinsic    FM     anisotropy,    assuming
$J_\mathrm{int}/J_\mathrm{FM}=0.4$.                          Evidently,
$\mathcal{K}_\mathrm{int}(T)$  assumes a finite  value for  an ordered
AFM phase, and disappears above the N\'eel temperature $T_N$.  If both
anisotropic contributions  are present, the  resulting total effective
anisotropy $\mathcal{K}_\mathrm{tot,FM}(T)$  is approximately given by
the        sum       of        $\mathcal{K}_\mathrm{int}(T)$       and
$\mathcal{K}_\mathrm{FM}(T)$, see the dot-dashed line (c). Finally, by
assuming  $K_\mathrm{FM}/J_\mathrm{FM}=-0.01$   the  dotted  line  (d)
refers to  the case  of an intrinsic  FM anisotropy favoring  the same
easy  axis  than   $K_\mathrm{int}$.  Therefore,  the  absolute  value
$|\mathcal{K}_\mathrm{tot,FM}(T)|$  may  be   either  reduced  (c)  or
enhanced (d)  by the interlayer exchange coupling.   We emphasize that
for the former  case the two anisotropies $\mathcal{K}_\mathrm{FM}(T)$
and  $\mathcal{K}_\mathrm{int}(T)$ compete,  resulting  possibly in  a
change  of sign  of $\mathcal{K}_\mathrm{tot,FM}(T)$,  and thus  in an
in-plane  magnetic   reorientation  of   the  FM  film   with  varying
temperature.
\begin{figure}
\includegraphics[width=5.1cm,height=7cm,bb=10    10    590   580,clip,
angle=-90]{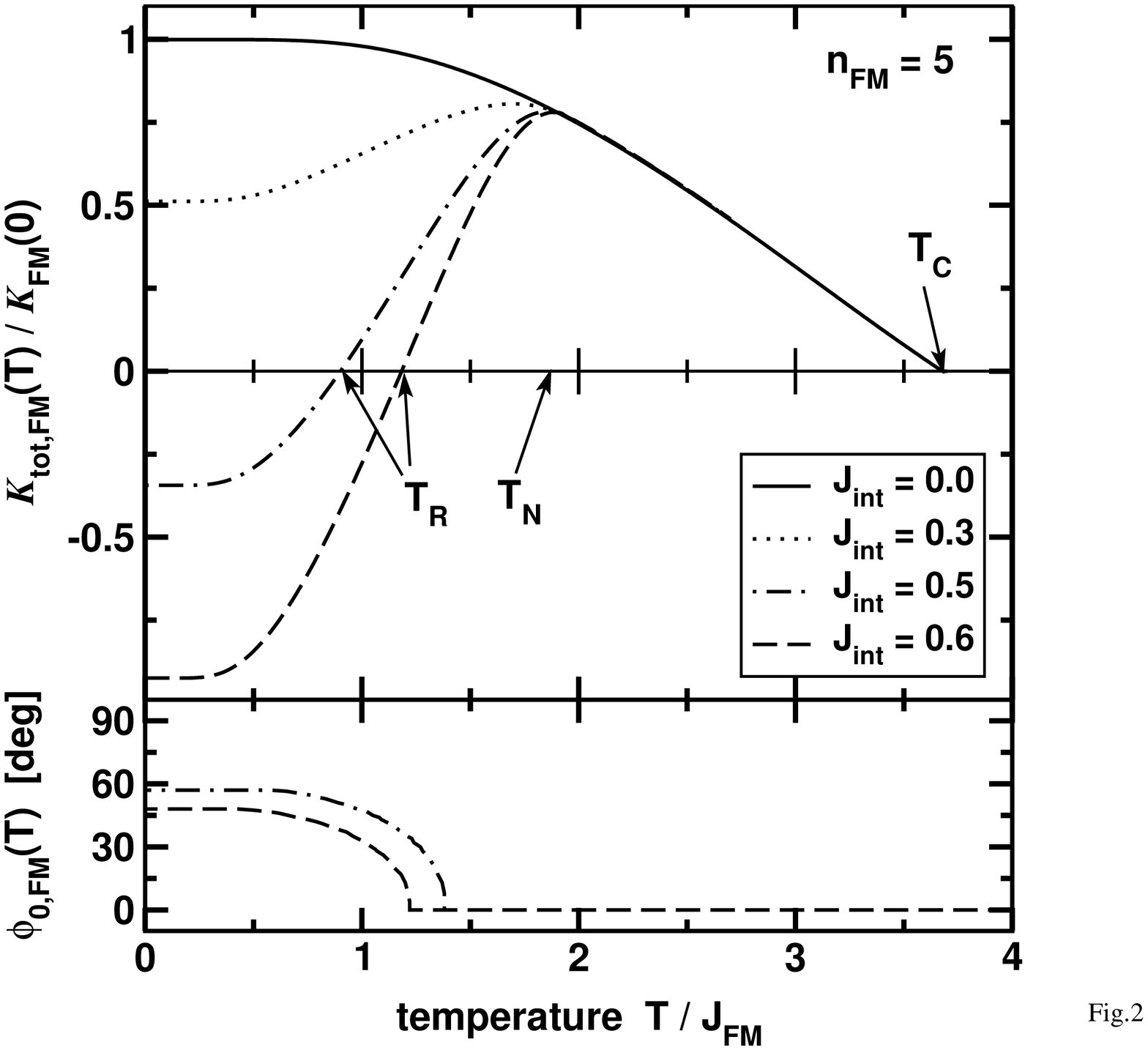}
\caption{ 
Total  effective  anisotropy  $\mathcal{K}_\mathrm{tot,FM}(T)$ per  FM
film  spin  as a  function  of  temperature  for different  interlayer
exchange couplings  $J_\mathrm{int}$, assuming $n_\mathrm{FM}=5$.  The
reorientation temperature  $T_R$ is defined  by the change of  sign of
$\mathcal{K}_\mathrm{tot,FM}(T)$.  In  addition the equilibrium angles
$\phi_\mathrm{0,FM}(T)$   of    the   FM   film    magnetization   for
$J_\mathrm{int}/J_\mathrm{FM}=0.5$                                  and
$J_\mathrm{int}/J_\mathrm{FM}=0.6$   are   displayed,   indicating   a
continuous  magnetic  reorientation  near $T_R$.   }  \vspace*{-0.5cm}
\end{figure}

Such  a magnetic  reorientation can  be  observed in  Fig.2, where  we
present $\mathcal{K}_\mathrm{tot,FM}(T)$  for different values  of the
interlayer  exchange coupling  $J_\mathrm{int}$  and for  the FM  film
thickness     $n_\mathrm{FM}=5$.      Furthermore,     Fig.3     shows
$\mathcal{K}_\mathrm{tot,FM}(T)$    for   different   $n_\mathrm{FM}$,
assuming  $J_\mathrm{int}/J_\mathrm{FM}=0.4$.  The  chosen  parameters
yield  $T_C>T_N$. As  mentioned, $K_\mathrm{FM}$  favors an  easy axis
collinear  to the AFM  magnetization. $\mathcal{K}_\mathrm{tot,FM}(T)$
changes  sign at  the  reorientation temperature  $T_R$  for a  strong
$J_\mathrm{int}$  or  for  a  small  $n_\mathrm{FM}$.  For  $T>T_R$  a
collinear,   and  for  $T<T_R$   preferably  an   orthogonal  magnetic
arrangement results.  This can be  seen from the  continuously varying
equilibrium FM angles $\phi_\mathrm{0,FM}$, which are also depicted in
Figs.2,3.    We   emphasize    that    the   orthogonal    arrangement
($\phi_\mathrm{0,FM}=\pi/2$) is not always realized. Rather, dependent
on the  interaction parameters  a canted magnetic  arrangement between
the  FM  and  the  AFM  subsystems  may  occur,  characterized  by  an
equilibrium angle $0<\phi_\mathrm{0,FM}<\pi/2$.  In this case the free
energy  $F(T,\phi_\mathrm{FM})$ as  a  function of  $\phi_\mathrm{FM}$
exhibits  four  minima  rather  than  two as  for  a  simple  uniaxial
anisotropy.\cite{ZSS01}  A two-fold  symmetry is  still  present.  For
very  strong $J_\mathrm{int}$  also  hysteresis effects  may occur  by
varying  $\phi_\mathrm{FM}$, accompanied  by sudden  jumps of  the AFM
spin  angles  $\phi_i$ (spin-flop-transition).\cite{HiM86,ScB98,CrC01}
The  AFM   film  exhibits  a   noncollinear,  spin-flop-like  magnetic
arrangement  for  $\phi_\mathrm{0,FM}>0$.   Furthermore,  an  in-plane
magnetic   reorientation  with   an  increasing   FM   film  thickness
$n_\mathrm{FM}$ for a constant temperature can be observed in Fig.3. A
small   value   for   $|\mathcal{K}_\mathrm{tot,FM}(T)|$  may   occur,
corresponding   to   a   very   soft  ferromagnet.    We   note   that
$\mathcal{K}_\mathrm{tot,FM}(T)$  does  not  depend  on  the  sign  of
$J_\mathrm{int}$,    consistent     with    the    estimate\cite{rem1}
$K_\mathrm{int}\propto-(J_\mathrm{int})^2/|J_\mathrm{AFM}|$.        The
disturbance of the AFM  spins in the spin-flop-phase decreases rapidly
with increasing distance from the interface.\cite{Jen01}
\begin{figure}
\includegraphics[width=5.1cm,height=7cm,bb=10    30    590   580,clip,
angle=-90]{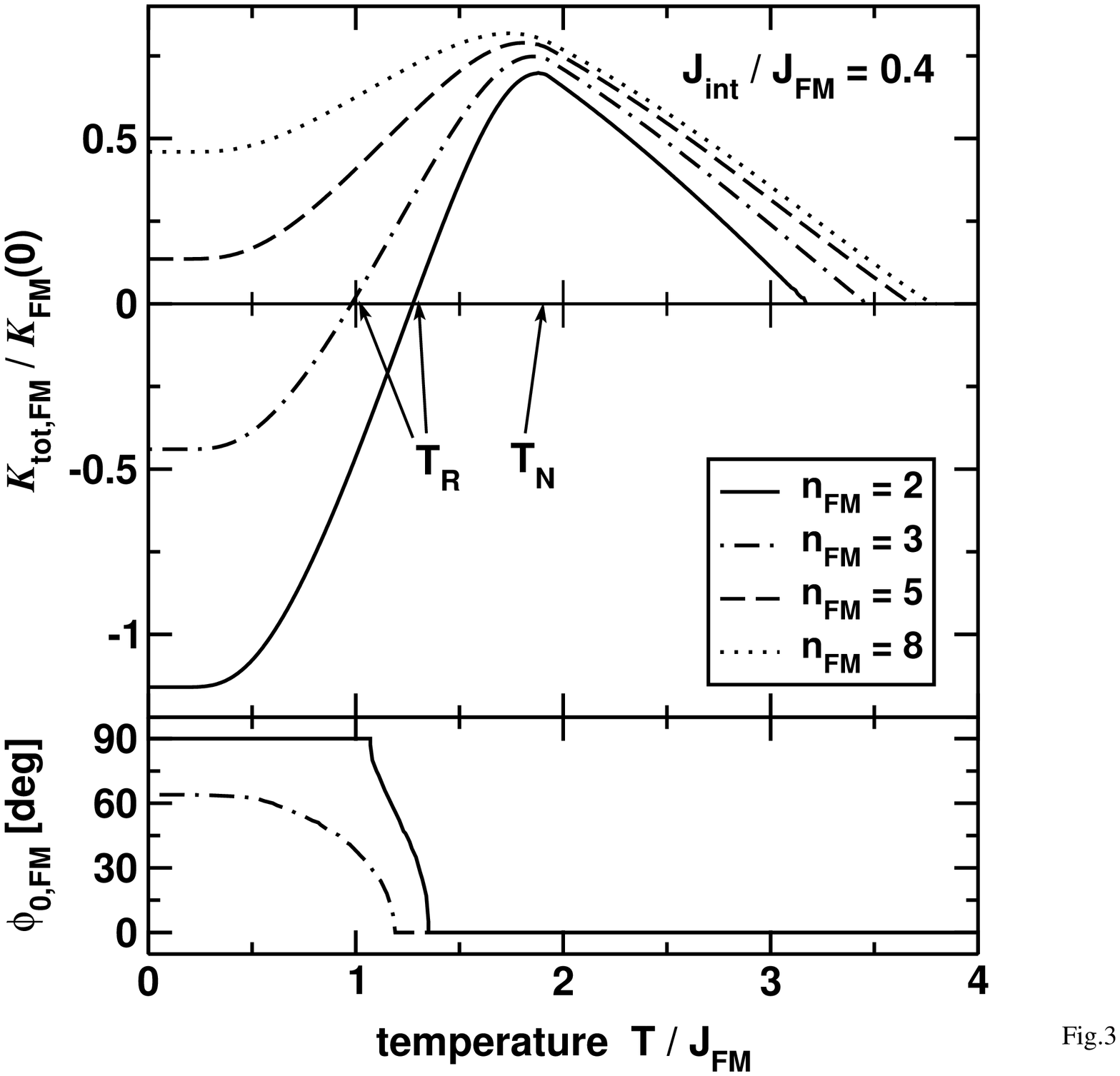}
\caption{ 
Total  effective  anisotropy  $\mathcal{K}_\mathrm{tot,FM}(T)$ per  FM
film  spin  as  a  function  of  temperature  for  different  FM  film
thicknesses                  $n_\mathrm{FM}$,                 assuming
$J_\mathrm{int}/J_\mathrm{FM}=0.4$.      The     equilibrium    angles
$\phi_\mathrm{0,FM}(T)$  of the  FM film  magnetization are  shown for
$n_\mathrm{FM}=2$ and $n_\mathrm{FM}=3$. The reorientation temperature
is  denoted by  $T_R$. Whereas  for $n_\mathrm{FM}=2$  a reorientation
between  the   collinear  and  orthogonal   magnetic  arrangements  is
obtained, for $n_\mathrm{FM}=3$ a canted arrangement is present at low
temperatures. } \vspace*{-0.5cm} \end{figure}

In  addition, we  present results  for the  case for  which  the Curie
temperature $T_C$  is smaller than the N\'eel  temperature $T_N$.  The
correspondent $\mathcal{K}_\mathrm{tot,FM}(T)$  is shown in  Fig.4 for
different interlayer exchange  couplings $J_\mathrm{int}$. By assuming
$n_\mathrm{FM}=1$ and  $J_\mathrm{AFM}/J_\mathrm{FM}=-0.75$, we obtain
$T_C/J_\mathrm{FM}=2.47$  and  $T_N/J_\mathrm{FM}=2.90$.  An  in-plane
magnetic reorientation of  the FM film close to  $T_C$ is obtained, if
$J_\mathrm{int}$  is  not  too  strong.   However, the  order  of  the
respective magnetic arrangements is  reversed with respect to the case
$T_C>T_N$.   In the  range $T_C<T<T_N$  a small  magnetic order  and a
small interface anisotropy $\mathcal{K}_\mathrm{int}(T)$ is induced in
the  FM film, resulting  in an  orthogonal magnetic  arrangement.  For
$T<T_C$  the   intrinsic  FM  anisotropy  $\mathcal{K}_\mathrm{FM}(T)$
becomes increasingly important and  may cause a magnetic reorientation
into  the   collinear  arrangement  with   a  decreasing  temperature.
Furthermore, an intermediate value of $J_\mathrm{int}$ can result in a
reentrant  magnetic behavior,  i.e.\  at lower  temperatures a  second
reorientation into  a canted  magnetic arrangement takes  place.  This
behavior    can    be   observed    from    the   equilibrium    angle
$\phi_\mathrm{0,FM}$,   which    is   also   shown    in   Fig.4   for
$J_\mathrm{int}/J_\mathrm{FM}=0.2$.    A   continuous   variation   of
$\phi_\mathrm{0,FM}$   is  obtained  for   low  temperatures,   and  a
discontinuous  one  close  to   $T_C$,  accompanied  by  a  hysteretic
behavior.   Note that the  results are  obtained under  the assumption
that  the magnetization  of the  FM film  stays always  parallel. This
assumption is questionable for temperatures $T_C<T<T_N$.

The strength of $J_\mathrm{int}$ is not well known.  It depends on the
material combination, the morphologyy,  and the presence of impurities
near the interface.  It  has been proposed to measure $J_\mathrm{int}$
by  applying   an  external  magnetic  field,   inducing  a  spin-flop
transition in the AFM subsystem.\cite{CrC01} However, to create such a
spin-flop transition the magnetic  field must possibly be very strong.
We propose  that $J_\mathrm{int}$ can  be determined by  measuring the
total anisotropy $\mathcal{K}_\mathrm{tot,FM}(T)$ of the FM film above
and below $T_N$, requiring $T_N<T_C$.
\begin{figure}
\includegraphics[width=5.1cm,height=7cm,bb=10    30    590   580,clip,
angle=-90]{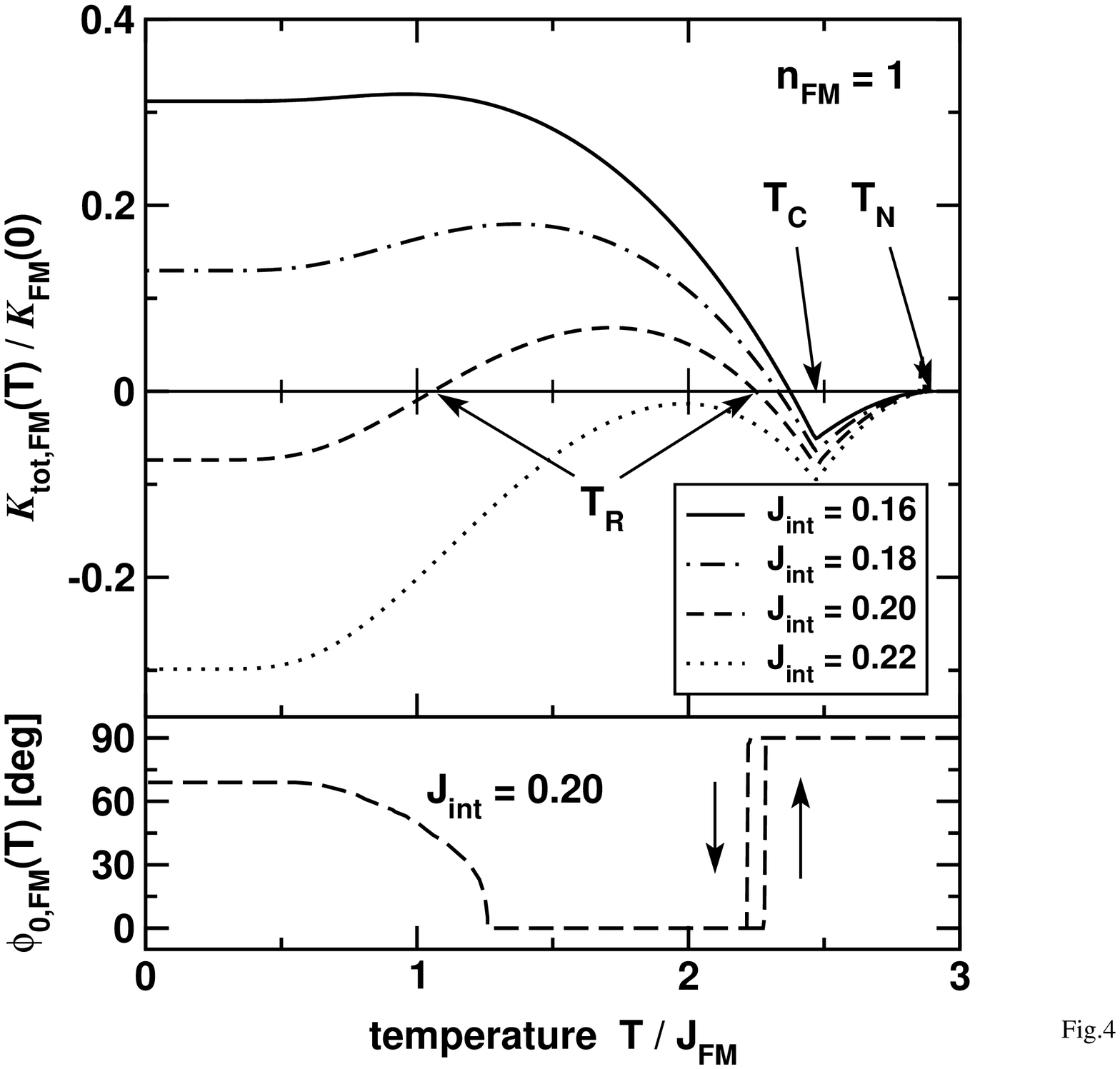}
\caption{ 
Total  effective  anisotropy  $\mathcal{K}_\mathrm{tot,FM}(T)$ per  FM
film  spin  as a  function  of  temperature  for different  interlayer
exchange  couplings $J_\mathrm{int}$,  assuming  $n_\mathrm{FM}=1$ and
$J_\mathrm{AFM}/J_\mathrm{FM}=-0.75$.   For these  values  one obtains
$T_C<T_N$.   The equilibrium angle  $\phi_\mathrm{0,FM}(T)$ of  the FM
film  magnetization is  shown  for $J_\mathrm{int}/J_\mathrm{FM}=0.2$.
For this value a reentrant magnetic behavior is obtained, indicated by
changes   of  sign   of   $\mathcal{K}_\mathrm{tot,FM}(T)$  near   the
reorientation temperatures  $T_R$, and  a hysteresis close  to $T_C$.}
\vspace*{-0.5cm} \end{figure}

In coupled FM-AFM systems also the exchange bias or the unidirectional
anisotropy is observed,\cite{NoS99,Kiw01} which is characterized by an
asymmetric  hysteresis loop.   Whereas  the origin  of this  important
quantity  is still  not  completely resolved,  the  occurrence of  the
exchange bias is  most likely caused by a  certain amount of interface
roughness,  defects, and  noncompensated  AFM spins,\cite{ScB98,TKB97}
accompanied    possibly   by    a    domain   phase    in   the    AFM
subsystem.\cite{Mal86}  In the  actual study  this phenomenon  has not
been adressed.

To  conclude, we  point out  the possibility  of an  in-plane magnetic
reorientation in coupled FM-AFM thin  films.  By application of a mean
field theory we have calculated the effective magnetic anisotropy, the
magnetic arrangement, and the equilibrium direction of the FM film for
such  systems.   The  interlayer  exchange  coupling  $J_\mathrm{int}$
causes  an interface  anisotropy  $\mathcal{K}_\mathrm{int}(T)$, which
adds to  the intrinsic anisotropy  $\mathcal{K}_\mathrm{FM}(T)$ of the
FM film,  and vanishes above the  N\'eel temperature $T_N$  of the AFM
system.   Depending on the  sign of  $\mathcal{K}_\mathrm{FM}(T)$, the
total anisotropy  $\mathcal{K}_\mathrm{tot,FM}(T)$ of the  FM film may
be enhanced as well as  be reduced, see Fig.1. For competing intrinsic
and interlayer  anisotropies a magnetic  reorientation of the  FM film
magnetization may  occur with increasing temperature $T$  or a varying
FM film thickness $n_\mathrm{FM}$, as  shown in Figs.2 -- 4.  The main
reason  for  the temperature  induced  in-plane  reorientation is  the
different  ordering  temperature  of  the two  subsystems,  causing  a
different  temperature dependency of  $\mathcal{K}_\mathrm{FM}(T)$ and
$\mathcal{K}_\mathrm{int}(T)$.  The  magnetizations of the  FM and AFM
films can be either collinear or orthogonal to each other. In addition
a canted  magnetic arrangement may  occur. Hence, the  assumption that
the  magnetic   structures  in  coupled  FM-AFM   systems  are  either
collinear\cite{Mal86}  or orthogonal\cite{Kiw01,Koo97}  is  not always
true.   The  noncollinear,   spin-flop-like  arrangement  of  the  AFM
spins\cite{HiM86,ScB98}  for  $\phi_\mathrm{0,FM}>0$ vanishes  rapidly
with  increasing  distance   from  the  FM-AFM  interface.   \\[0.2cm]
\hspace*{0.5cm} Numerous  discussions with M.\  Alouani are gratefully
acknowledged. P.J.J.\ acknowledges financial support from the European
Union TMR 'Interface Magnetism', grant No.\ ERBFMXCT 96 - 0089.


\begin{thebibliography}{99}
%
\bibitem{NoS99} For  a review,  see J. Nogu\'es  and I.  K.  Schuller,
J. Magn. Magn. Mater. \textbf{192}, 203 (1999).
%
\bibitem{Kiw01}  M.  Kiwi, J.  Magn.  Magn.  Mater. \textbf{234},  584
(2001).
%
\bibitem{Koo97}  N.  C. Koon,  Phys.   Rev.   Lett. \textbf{78},  4865
(1997).
%
\bibitem{HiM86}  L.  L.  Hinchey  and   D.  L.  Mills,  Phys.  Rev.  B
\textbf{34}, 1689  (1986); R. W.  Wang and D.  L. Mills, Phys.  Rev. B
\textbf{50}, 3931 (1994).
%
\bibitem{ScB98} T.  C. Schulthess and  W. H. Butler, Phys.  Rev. Lett.
\textbf{81}, 4516 (1998).
%
\bibitem{rem1}  Consider  a  FM  and  an AFM  layer  with  a  mutually
orthogonal magnetization coupled by  $J_\mathrm{int}$. If the spins of
the two AFM sublattices  are rotated clockwise and counterclockwise by
the angle  $\phi$ into the  direction of the  FM, the energy  per unit
cell                   is                   given                   by
$E(\phi)=-2\,|J_\mathrm{AFM}|\,\cos(2\,\phi)-J_\mathrm{int}\,\sin\phi$.
Minimization  with  respect to  $\phi$  yields  the equilibrium  angle
$\phi_0=J_\mathrm{int}/(8\,|J_\mathrm{AFM}|)$.  The  energy difference
between  the disturbed  and undisturbed  AFM arrangement  is  given by
$\Delta E= E(\phi_0)-E(0)=-(J_\mathrm{int})^2/(16\,|J_\mathrm{AFM}|)$.
%
\bibitem{MNL98}  T. J.  Moran, J.  Nogu\'es, D.  Ledermann, and  I. K.
Schuller,  Appl.  Phys.  Lett.  \textbf{72},  617  (1998);  Y.  Ijiri,
J.  A. Borchers,  R. W.  Erwin, S.  H. Lee,  P. J.  van der  Zaag, and
R. M. Wolf, Phys. Rev. Lett. \textbf{80}, 608 (1998).
%
\bibitem{Jen01}  P. J.  Jensen,  Appl. Phys.  Lett. \textbf{78},  2190
(2001).
%
\bibitem{CaC66}  H. B.   Callen and  E. R.   Callen, J.   Phys.  Chem.
Solids \textbf{27}, 1271 (1966).
%
\bibitem{JeB98} For a review, see P. J. Jensen and K. H. Bennemann, in
\textit{Magnetism and Electronic Correlations in Local-Moment Systems:
Rare Earth Elements and Compounds,} edited by M. Donath, P. A. Dowben,
and W. Nolting (World Scientific, Singapore, 1998), pp. 113 --140.
%
\bibitem{noncom} A  noncompensated interface can  simply be considered
by assuming different interlayer couplings $J_\mathrm{int}^{I,II}$ for
the two AFM sublattices across the interface.\cite{CrC01}
%
\bibitem{rem2} The assumed spin quantum number $S=1$ is the lowest one
to take  into account  a second order  single-ion anisotropy.  With an
enhanced computational  effort also larger $S$ or  classical spins can
be considered as well.
%
\bibitem{HeB94}   See,   for   example,   \textit{Ultrathin   Magnetic
Structures I + II,} edited by B. Heinrich and J. A. C. Bland (Springer
Verlag, Berlin, 1994).
%
\bibitem{Ogu55}  T.  Oguchi,  Progr.  Theor.  Phys.  \textbf{13},  148
(1955); J.  S. Smart, \textit{Effective  Field Theories  of Magnetism}
(Saunders, 1966).
%
\bibitem{ZSS01}  W.   Zhu,  L.  Seve,  R.  Sears,   B.  Sinkovic,  and
S. S. P. Parkin, Phys.  Rev.  Lett. \textbf{86}, 5389 (2001).
%
\bibitem{CrC01} N. Cramer and R.  E. Camley, Phys. Rev. B \textbf{63},
60404 (2001).
%
\bibitem{TKB97} K. Takano, R. H.  Kodama, A. E. Berkowitz, W. Cao, and
G.  Thomas, Phys.  Rev.   Lett. \textbf{79},  1130 (1997);  U.  Nowak,
R. W. Chantrell, and E.  C. Kennedy, Phys. Rev. Lett. \textbf{84}, 163
(2000).
%
\bibitem{Mal86}  A. P.  Malozemoff, Phys. Rev.  B \textbf{34},
1853 (1986); \textit{ibid.} \textbf{35}, 3679 (1987); 
D. Mauri, H. C. Siegmann, P. S. Bagus, and E. Kay, 
J. Appl. Phys. \textbf{62}, 3047 (1987). 
%
\end{thebibliography}
\end{document}